\pdfoutput=1
\documentclass{PoS}
\usepackage{graphicx}
\usepackage{subfigure}
\usepackage{amsmath}

\newcommand{\skipthis}[1]{}

\title{SU(3) Deconfinement in (2+1)d 
 from Twisted Boundary Conditions and Self-Duality}

\ShortTitle{SU(3) Deconfinement in (2+1)d}

\author{\speaker{Nils Strodthoff}\\
        Institut f\"ur Kernphysik, Technische Universit\"at Darmstadt, D-64289 Darmstadt, Germany\\
        E-mail: \email{nstrodt@crunch.ikp.physik.tu-darmstadt.de}}
\author{Sam R. Edwards\\
        Institut f\"ur Kernphysik, Technische Universit\"at Darmstadt, D-64289 Darmstadt, Germany\\
        E-mail: \email{edwards@crunch.ikp.physik.tu-darmstadt.de}}
\author{Lorenz von Smekal\\
        Institut f\"ur Kernphysik, Technische Universit\"at Darmstadt, D-64289 Darmstadt, Germany\\
        E-mail: \email{lorenz.smekal@physik.tu-darmstadt.de}}

\abstract{We study the pure SU(3) gauge theory in 2+1
  dimensions on the lattice using 't Hooft's twisted boundary
  conditions to force non-vanishing center flux through the finite
  volume. In this way we measure the free energy of spacelike center
  vortices as an order parameter for the deconfinement transition. The
  transition is of $2^\mathrm{nd}$ order 
  in the universality class of the 2$d$ 3-state Potts
  model, which is self-dual. This self-duality can be observed
  directly in the SU(3) gauge theory, and it can be exploited to extract
  critical couplings with high precision in rather small volumes. We
  furthermore obtain estimates for critical exponents and the critical
  temperature in units of the dimensionful continuum coupling. Finally,
  we also apply our methods to the (2+1)$d$ SU(4) gauge theory which
  was previously found to have a weak $1^\mathrm{st}$ order
  transition. We nevertheless observe at least approximate $q=4$ Potts
  scaling at length scales corresponding to the lattice sizes used
  in our simulations.} 

\FullConference{The XXVIII International Symposium on Lattice Field Theory, Lattice2010\\
		June 14-19, 2010\\
		Villasimius, Italy}

\begin{document}

\vspace*{-1.2cm}

\section{Introduction}

\vspace{-.2cm}

The motivation of our study is to see how much we can learn
about the deconfinement transitions in pure SU($N$) gauge theories at
finite temperature  from universality and scaling. As these concepts
require  second order phase transitions, in 3+1 dimensions we are only
left with SU(2), where spatial center vortex sheets share their universal
behavior with interfaces in the 3$d$ Ising model. This has been
studied in \cite{deForcrand:2001nd}. In 2+1 dimensions on the other
hand, we can apply analogous methods to both SU(2) and SU(3) which
then exhibit a $2^\mathrm{nd}$ order deconfinement transition.
The latter is in the universality class of the 2$d$ Ising model for
which by far the largest pool of exact results is available.
This was exploited in \cite{Edwards:2009qw} for a high precision
determination of critical couplings and temperature, and in
\cite{Strodthoff:2010} for accurate finite-size scaling, a
reflection of self-duality and a precision determination of the
behavior of the continuum sting tension and its dual around the phase
transition. Here, we report on first results from applying the same
methods to SU(3) and gradually also to SU(4) in 2+1 dimensions.

For SU(3), the dimensionally reduced spin model with the
same global symmetry and the universal properties of the $Z_3$
center-symmetry breaking deconfinement transition is the
3-state Potts model. For SU(4), the $Z_4$ center symmetry 
alone does not uniquely specify the effective spin model to describe
the dynamics of Polyakov loops. SU(4) is a rank-three group and has
three fundamental representations, $4$, $\bar 4$ and $6$. So even the
simplest effective Polyakov-loop model will consist of two distinct
real terms, with nearest neighbor couplings between loops in $4$/$\bar
4$ representations and between loops in the $6$ representation
\cite{Wozar:2006fi}. Depending on the relative weight between the two,
the corresponding spin model could be any of the $Z_4$-symmetric
Ashkin-Teller models with three energy levels per link and
continuously varying critical exponents between the $q=4$ Potts model
class and that of the planar or vector Potts model which corresponds
to two non-interacting Ising models in this case. Even though the
more recent studies of the (2+1)$d$ SU(4)
gauge theory \cite{deForcrand:2003wa,Holland:2007ar,Liddle:2008kk}
indicate that the transition is weakly $1^\mathrm{st}$ order, we do find, at
least approximately at the length scales corresponding to our spatial
lattice volumes, a universal scaling which seems closest to the standard
$q=4$ Potts case. Because this might not seem very natural, it would
be interesting to explain that. 

One special feature of the $q$-state Potts models in 2 dimensions is
that they are self-dual for all $q$, no matter whether they have
$1^\mathrm{st}$ ($q> 4$) or $2^\mathrm{nd} $ ($q\le 4$) order
transitions. With a $2^\mathrm{nd} $ order transition and scaling,
this self-duality is reflected in the (2+1)$d$ gauge theory: the
spatial center-vortex free energies are mirror images around
criticality of those of the confining electric fluxes
\cite{Strodthoff:2010}. Here we verify this explicitly for SU(3), and
show how it can be used to remove the leading (universal) 
finite-size corrections in the determination of critical couplings
from finite volume extrapolations.   

\vspace{-.4cm}

\section{Concepts and Methods}

\vspace{-.2cm}

\paragraph{'t Hooft's twisted boundary conditions, center vortices and
  electric fluxes:}
In a theory without matter fields where the gauge fields represent the
center of the gauge group trivially, the boundary conditions in a
finite Euclidean $1/T\times L^d$ volume are only fixed up to center elements
giving rise to $N^d$ gauge-inequivalent boundary conditions for a pure
SU($N$) gauge theory in $d$+1 dimensions. These twisted boundary
conditions \cite{'tHooft:1979uj} can be classified either as magnetic
twists defined in purely spatial planes or as temporal twists in the
planes oriented along the Euclidean time direction. The latter
are labeled by a vector $\vec{k}\in{Z}_N^d$. In the
following we will only consider temporal twists because the magnetic
twists are irrelevant for the deconfinement transition.

Temporal twist introduces spatial center vortices whose free energies
provide order parameters for the deconfinement transition.
These vortex free energies $F_k$ (per temperature  $T$) are defined
as ratios $R_k(\vec k) \equiv Z_k(\vec{k})/Z_k(\vec{0}) =
e^{-F_k(\vec{k})}$ of partition functions $Z_k(\vec k)$ with temporal
twist $\vec k$ over the periodic ensemble $Z_k(\vec 0)$. Analogously,
one defines the electric flux free energies $F_e$ via $R_e(\vec
e)\equiv Z_e(\vec{e})/Z_e(\vec{0}) = e^{-F_e(\vec{e})}$. These
describe gauge-invariant color-singlet free energies of static
fundamental charges at some point $\vec x $ with mirror anti-charges in a
neighboring volume at $\vec x + L\vec e$ along the direction of the
flux $\vec e$ relative to the {\em no-flux} ensemble $Z_e(\vec 0) =
\sum_{\vec k} Z_k(\vec k)$, 
which is an enlarged ensemble corresponding to fluctuating temporal
twists, see \cite{deForcrand:2001nd,Strodthoff:2010}.
The electric flux and spatial center vortex partition functions are
related by a $d$-dimensional ${Z}_N$ Fourier transform, 
\begin{equation} 
\label{eq:relzezk}
R_e(\vec{e})  \, =\, \frac{1}{N}\, \big\langle \mbox{tr}\big( P(\vec
x) P^\dagger (\vec x+ \vec e L) \big) \big\rangle_{\mbox{\scriptsize no-flux}}\,
= \, \big({\textstyle \sum_{\vec{k}}}\; e^{2\pi\text{i}\, \vec{e}\cdot
  \vec{k} /N}\,R_k(\vec{k})\big)/{\textstyle \sum_{\vec{k}}} \;
R_k(\vec{k}) \; . 
\end{equation}

\skipthis{\\The relevance of center vortices for the deconfinement
  transition stems from the following intuitive picture for the
  deconfinement transition: Spacelike vortices (corresponding to
  temporal twists) can link with timelike Wilson loops. These pierced
  Wilson loops then get disordered and acquire a phase. At
  temperatures below the critical temperature the vortices can spread
  to lower their free energy forming so-called thick vortices which
  then give rise to an area law falloff for timelike Wilson loops. At
  high temperatures timelike vortices get squeezed and are essentially
  suppressed. Timelike vortices can spread at any temperature giving 
  rise to an area law falloff for spacelike Wilson loops at any
  temperature. This explains why they are irrelevant for the phase
  transition.\\  
  As order parameters we have the center vortex free energy $F_k\sim
  \tilde{\sigma}L^{d-1}$ with the dual string tension
  $\tilde{\sigma}>0$ for $T>T_c$. A dual order parameter is the
  electric flux free energy $F_e\sim \sigma L/T$ with string tension
  $\sigma>0$ for $T<T_c$.} 

\vspace{-.2cm}

\paragraph{Universality and self-duality:}
By the Svetitsky-Yaffe conjecture, 
a $d+1$ dimensional gauge theory with second order deconfinement
transition has the same universal properties as a $d$ dimensional spin
model with the same global $Z_N$ symmetry \cite{Svetitsky:1982gs}. 
The Polyakov loop correlators of the gauge theory near criticality
behave in the same way as those of spins in the spin model. Spatial
center vortices correspond to spin interfaces, which are frustrations
where the coupling of adjacent spins favors cyclically shifted spin
states rather than parallel ones for the usual ferromagnetic
couplings. Consequently, the center-vortex free energies $F_k$ show
the universal behavior of interface free energies. For the 2+1
dimensional SU(3) gauge theory the corresponding spin model is the
2$d$ 3-state Potts model. Like all $q$-state Potts models in $2$
dimensions, it is self-dual. 

Kramers-Wannier duality is of course a very well-known concept from statistical
physics \cite{Savit:1979ny}. It provides exact maps between the spin
systems and their dual theories in terms of disorder variables on the 
dual lattice. In 2 dimensions, just as for the Ising model ($q=2$)
these dual theories of the $q$-state Potts models are again $q$-state Potts
models, but at a dual temperature $\widetilde T$ which is swapped
around criticality at $T_c$ as compared to the original model. Duality
transformations in a finite volume do not preserve boundary
conditions, however. Periodic boundary conditions on one side
generally correspond to fluctuating boundary conditions on the other
\cite{Gruber:1977}. This was explicitly demonstrated for the duality
between the 3$d$ Ising and the $Z_2$-gauge model in
\cite{Caselle:2001im}. The exact finite-volume duality transformation
for the $2d$ $q$-state Potts models is given in
\cite{Strodthoff:2010}. Here it suffices to note that its structure is
precisely that in (\ref{eq:relzezk}) with $q=N$. It expresses the partition
function of the dual $q$-state Potts model with certain set boundary
conditions at a temperature $\widetilde T$ as a 2$d$ $Z_q$-Fourier
transform over Potts models with all possibilities of cyclically
shifted boundary conditions at temperature $T$. In the (2+1)$d$ gauge
theory, the temperature is the same on both sides of the $Z_N$-Fourier
transform (\ref{eq:relzezk}). But within the universal scaling
window around a $2^\mathrm{nd}$ order phase transition, as a
consequence of the self-duality of the spin model, the free energies
of spatial center vortices and those of the confining electric fluxes
are mirror images of one another around $T_c$.

\vspace{-.2cm}

\paragraph{Numerical procedure:}
To implement 't Hooft's twists on the lattice we
multiply a stack of plaquettes by the corresponding center element 
$z\in Z_N$ so as to fix the corresponding amount of center flux
through the planes with twisted boundary conditions. For the temporal
$\vec k$-twists this introduces thin spatial center vortices
perpendicular to $\vec k$ which separate regions where fundamental
Polyakov loops differ by the center phase $z = e^{2\pi i \,
  k/N}$. They are thus like spin interfaces in the Potts model.  

We use the snake algorithm \cite{deForcrand:2000fi} to compute the
ratios $R_k(\vec k) = Z_k(\vec{k})/Z_k(\vec 0)$ via products of ratios of
partition functions which differ by only one phase rotated plaquette
to successively form a vortex that wraps around the spatial volume
perpendicular to $\vec k$. We use the standard Wilson gauge action
and a heatbath algorithm supplemented by a number of overrelaxation
steps. SU($N$) updates are implemented as updates of SU(2) subgroups
in the usual way \cite{Cabibbo:1982zn}. 

\vspace{-.4cm}

\section{Results}

\vspace{-.2cm}

\paragraph{Self-duality in SU(3):}
As for $SU(2)$ in 2+1 dimensions \cite{Strodthoff:2010} the
self-duality of the 3-state Potts model is reflected in SU(3). 
The spatial center vortex and electric flux partition
functions are related by a $Z_3$-Fourier transform as in 
(\ref{eq:relzezk}) whose structure is the same as that of
the finite volume duality transformation of the Potts model. Its
self-duality therefore implies that electric fluxes on one side of the
phase transition should resemble center vortex ensembles on the other.

\begin{figure*}
\centering
\subfigure{
\includegraphics[width=0.47\textwidth]{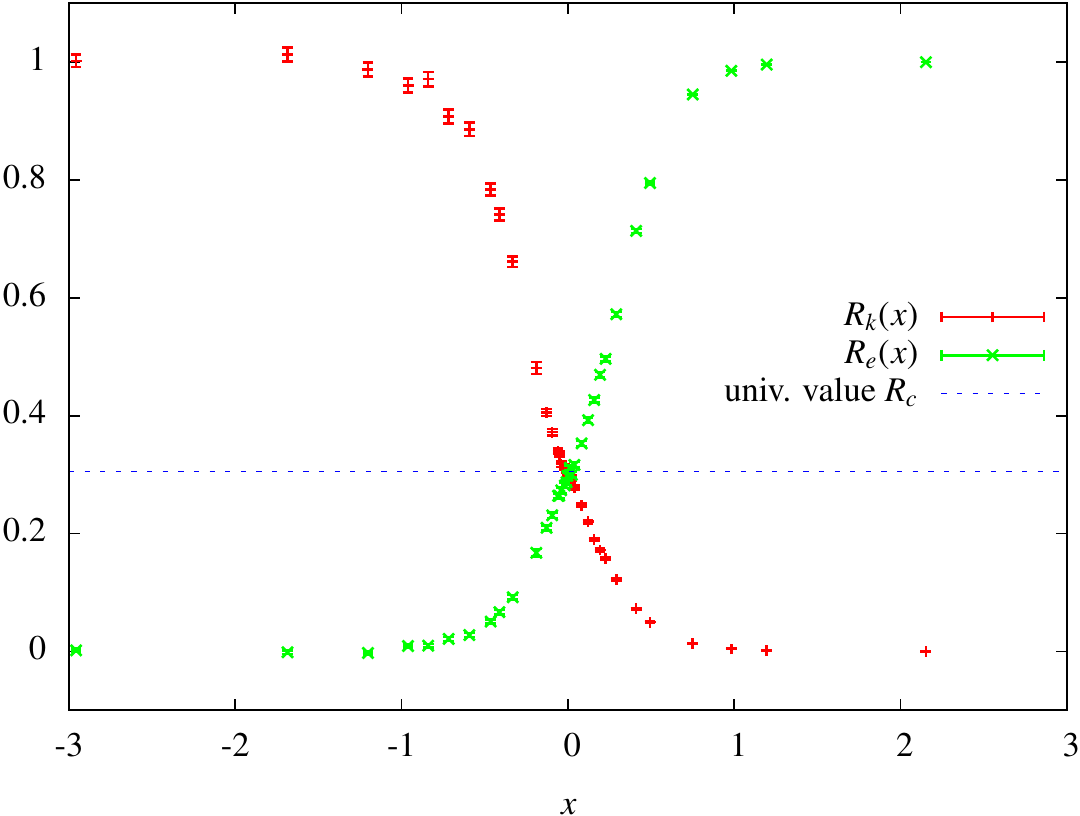}
\label{fig:subfigsu3no1}
}
\subfigure{
\includegraphics[width=0.47\textwidth]{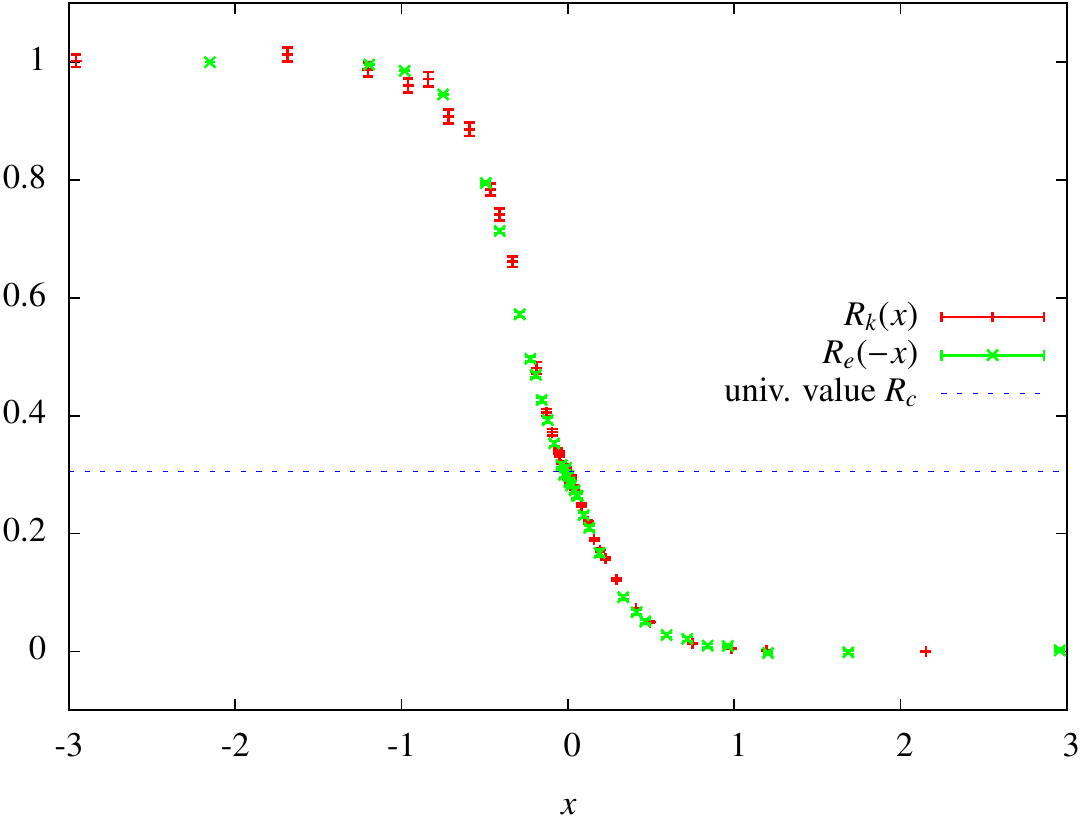}
\label{fig:subfigsu3no2}
}
\vspace{-.2cm}
\label{fig:su3selfduality}
\caption{Self-duality in SU(3) ($N_s=24$, $N_t=2$)
\vspace{-.6cm} }
\label{fig:sdsu3}
\end{figure*}

Swapping the temperature $\widetilde T \leftrightarrow T$ in the spin
model near criticality amounts to $x \leftrightarrow -x$ where $x =
\pm L/\xi_\pm$ is the finite-size scaling variable given by the ratio
of the finite size $L$ over the correlation lengths $\xi_\pm = f_\pm
(\pm t)^{-\nu} $, with $\xi_+$ for $x>0$ and $t > 0 $
above $T_c$, and $\xi_-$ for $x,\,t<0$, below.  For the
gauge theory we use  $x=\pm T_c L\,(\pm t)^\nu\propto L/\xi^{\pm} $ and
control the reduced temperature $t=T/T_c-1$ by changing the lattice
coupling. Within the universal scaling window, as functions of $x$,
we should then find that $R_k(x) \, = \, R_e(-x) $ for matching
pairs of twist $\vec k$ and flux $\vec e$. This is indeed the case 
also for SU(3), as demonstrated in Fig.~\ref{fig:sdsu3}, where we
compare the ratios of partition functions $R_k$ and $R_e$ for one unit
of temporal twist and one unit of electric flux, {\it i.e.}, both $\vec k$
and $\vec e$ either $(1,0)$ or $(0,1)$, over the phase transition.

\begin{figure*}
\centering
\subfigure{
\includegraphics[width=0.47\textwidth]{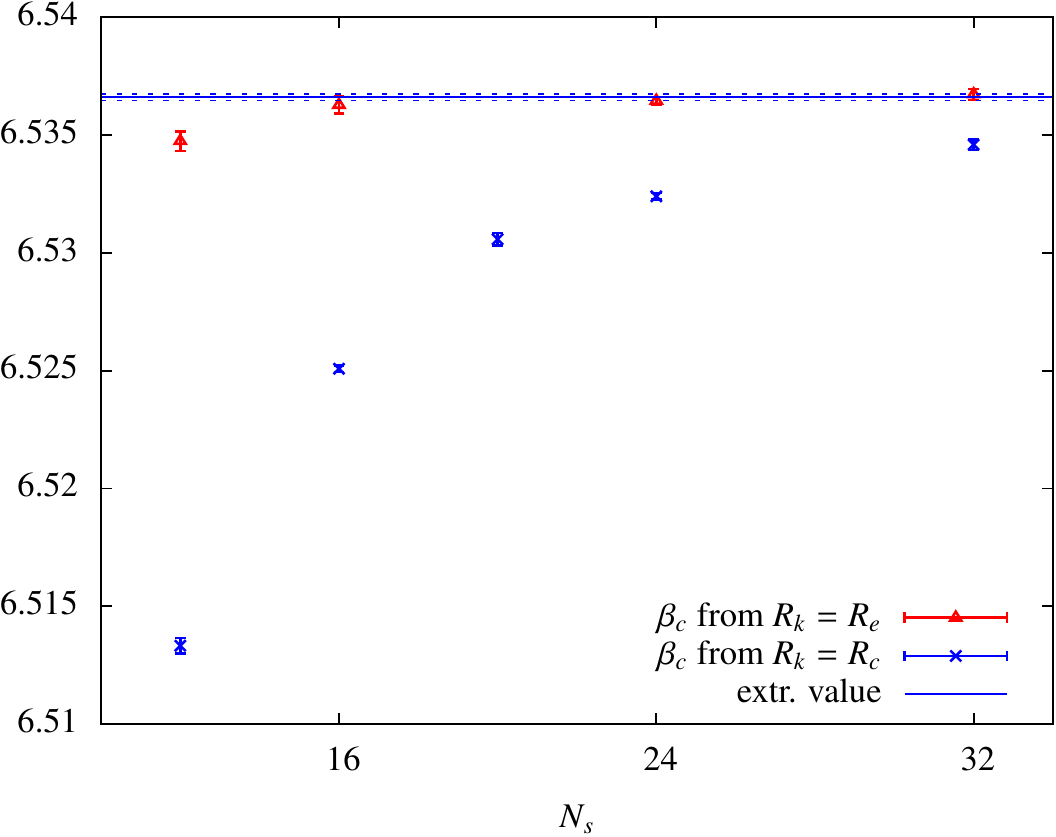}
\label{fig:su2convergence4compare}

}
\subfigure{
\includegraphics[width=0.47\textwidth]{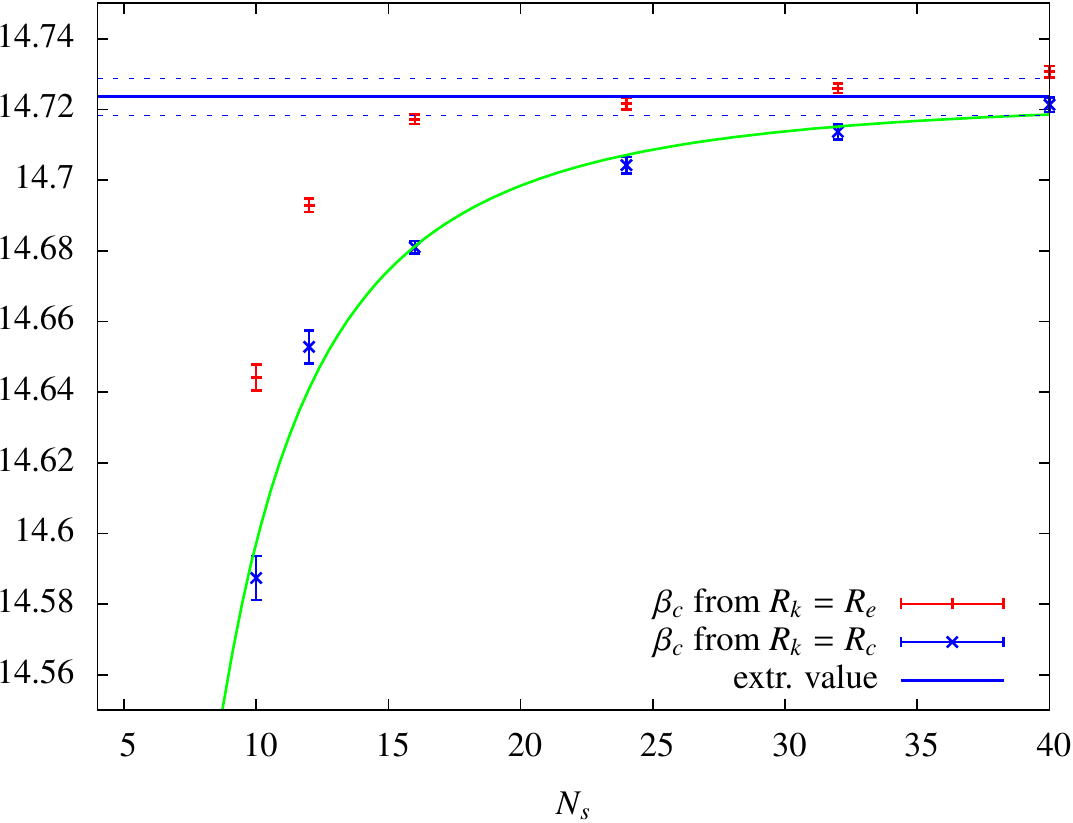}
\label{fig:su3convergence4compare}
}
\vspace{-.2cm}
\label{fig:sunconvergence4compare}
\caption{Convergence to $\beta_{c}$ ($N_t=4$) from $R_k(\beta) =
  R_e(\beta)$ and $R_k(\beta) = R_c$ in SU(2) (left) and SU(3)
  (right).}
\end{figure*}

\vspace{-.2cm}

\paragraph{Critical couplings:} There is a long history of methods to
extract critical couplings or temperatures from simulations in finite
volumes, going back to using pairwise intersections of Binder
cummulants on sucessively larger lattices \cite{Binder:1981sa}. 
Hasenbusch later demonstrated that the ratios of
partition functions with different boundary conditions 
could be used in the same way to obtain a much more rapid convergence
with very good estimates already from rather small lattices
\cite{Hasenbusch:1993wn}.  At criticality, these ratios tend to
universal values $0<R_c<1$ in the thermodynamic limit. In
\cite{Edwards:2009qw} it was therefore shown how to obtain critical
couplings for gauge theories from intersecting the ratios $R_k$ of finite
volume partition functions with these universal fixed points, once 
their values are known. For (2+1)$d$ SU(2) this led to an even faster
convergence than their pairwise intersections. For the 2$d$ Potts models
with $2^\mathrm{nd}$ order transition, {\it i.e.}, for $q=2,3$ and 4,
the universal numbers $R_c^{(m,n)}$ have been obtained exactly,
in terms of Jacobi theta functions, for all cyclic boundary conditions
(with $m,n = 0,1,\dots q-1$) in  \cite{PhysRevB.38.565}. For $q=3$
on a symmetric lattice they are,   
\begin{equation} 
R_c^{(1,0)}=R_c^{(2,0)} = 0.30499982\dots \; , \;\; \mbox{and} \;\;\; 
R_c^{(1,1)}=R_c^{(2,2)} = R_c^{(1,2)} =  0.19500018\dots \; . 
\end{equation}
Using a finite-size scaling ansatz for the vortex ensemble ratios
$R_k$ around criticality of the form  
\begin{equation}
R_k(\beta)=R_c+ b \left(\beta -\beta_{c} \right)N_s^{1/\nu}+c
N_s^{-\omega}+\cdots, \label{fssans}  
\end{equation} 
we define pseudo-critical couplings $\beta_c(N_s,N_t)$ in a finite
volume by requiring that $R_k(\beta) = R_c$, 
\begin{equation}
\beta_c(N_s,N_t)=\beta_{c}(N_t)-(c/b)\, N_s^{-(\omega+1/\nu)} + \cdots .
\end{equation}
These extrapolate to $\beta_{c}(N_t)$ from large spatial lattice sizes
$N_s$ at fixed numbers of time slices $N_t$. As a byproduct this
method gives numerical estimates of the correction to scaling
exponent $\omega$.

\begin{floatingtable}[r]
\centering
    \begin{tabular}{|c|c|c|c|}
\hline  $N_t$&$\beta_c$($R_k = R_e$)  & $\beta_c$ ($R_k = R_c$) &Lit.\\ 
\hline  2& 8.15309(11)& 8.15297(57) & 8.1489(31)$^\dagger$ \\ 
\hline  4& 14.7262(9)& 14.7194(45)  & 14.717(17)$^\dagger$ \\
\hline  6& 21.357(25)& - & 21.34(4) $^\ddagger$\\
\hline  8& 27.84(12)& - &  -\\ 
\hline 
\end{tabular}
\caption{SU(3) critical couplings from self-duality (weighted means), intersection with the universal value (extrapolated), and literature values from
  $^\dagger$\cite{Liddle:2008kk},  $^\ddagger$\cite{Engels:1996dz}.}
\label{tab1}
\end{floatingtable}

With self-duality, however, there is a yet more efficient method to
determine $\beta_{c}$ \cite{Strodthoff:2010}. This is based on the
simple observation that one must then have  $R_e(\beta ) = R_k(\beta)$
for like $\vec e$ and $\vec k$ at $\beta=\beta_c$. In fact, one can
easily convince oneself that with self-duality,
\[ 
R_e(\beta)=R_c -  b \left(\beta -\beta_{c} \right)N_s^{1/\nu}+c
N_s^{-\omega}+\cdots,  
\] 
 with the same coefficients $b$ and $c$ as in~(\ref{fssans}).

Therefore, the leading finite-size corrections to $\beta_{c}$ when defined
by $R_e = R_k$ cancel. At criticality, $R_e(\beta_c) = R_k(\beta_c)
= R_c + c N_s^{-\omega} + \cdots$, so the leading corrections only
move the intersection point 
vertically without shifting the so defined critical coupling. 
The gain is illustrated for both SU(2) and SU(3) in
Fig.~\ref{fig:sunconvergence4compare}.  
When intersecting  $R_e = R_k$, we form weighted means 
from sufficiently large aspect ratios $N_s/N_t$ where we assume that the
estimates have converged within errors. These are
compared to the extrapolated values 
from intersecting $R_k(\beta)$  with the universal value 
$R_c$ in Tab.~\ref{tab1}. 


\vspace*{-.2cm}

\paragraph{Critical temperature and correlation length exponent
  $\nu$:} In 2+1 dimensions the critical couplings grow linearly with $N_t$ to
leading order at large $N_t$. The slope is given by the
critical temperature in units of the dimensionful continuum coupling,
 $\beta_c(N_t)/(2N_c)  =  (T_c/g^2_3) \, N_t + \cdots $. 

From our values for $N_t=4$, 6 and 8 we then obtain
$T_c/{g_3^2}=0.5475(3)$ corresponding to
${T_c}/{\sqrt{\sigma}}=0.9938(9)$ with a zero temperature string
tension $\sqrt{\sigma}/g_3^2 = 0.5509(4) $ from a weighted average of
the four values in \cite{Bringoltz:2006zg}. This is consistent with 
$T_c/{\sqrt{\sigma}}=0.9994(40)$ from \cite{Liddle:2008kk}. 

Moreover, because the spatial center vortex free energies $F_k$ 
for sufficiently large $L$ depend only on $L^{1/\nu} t$, and $t
\propto (\beta -\beta_c)$, when
expanding $F_k(\beta ) = -\ln R_c + d(N_s) (\beta-\beta_c)+ \cdots $,
we can expect the slope at  $\beta_c$ to behave as $d(N_s)\sim N_s^{1/\nu}$.
The result from fitting our slopes for SU(3) with $N_t=4$ then gives
$\nu=0.82(4)$ as compared to $\nu=5/6\approx 0.833$ for the 2$d$ 3-state
Potts model.

\begin{figure*}
\centering
\subfigure{
\includegraphics[width=0.47\textwidth]{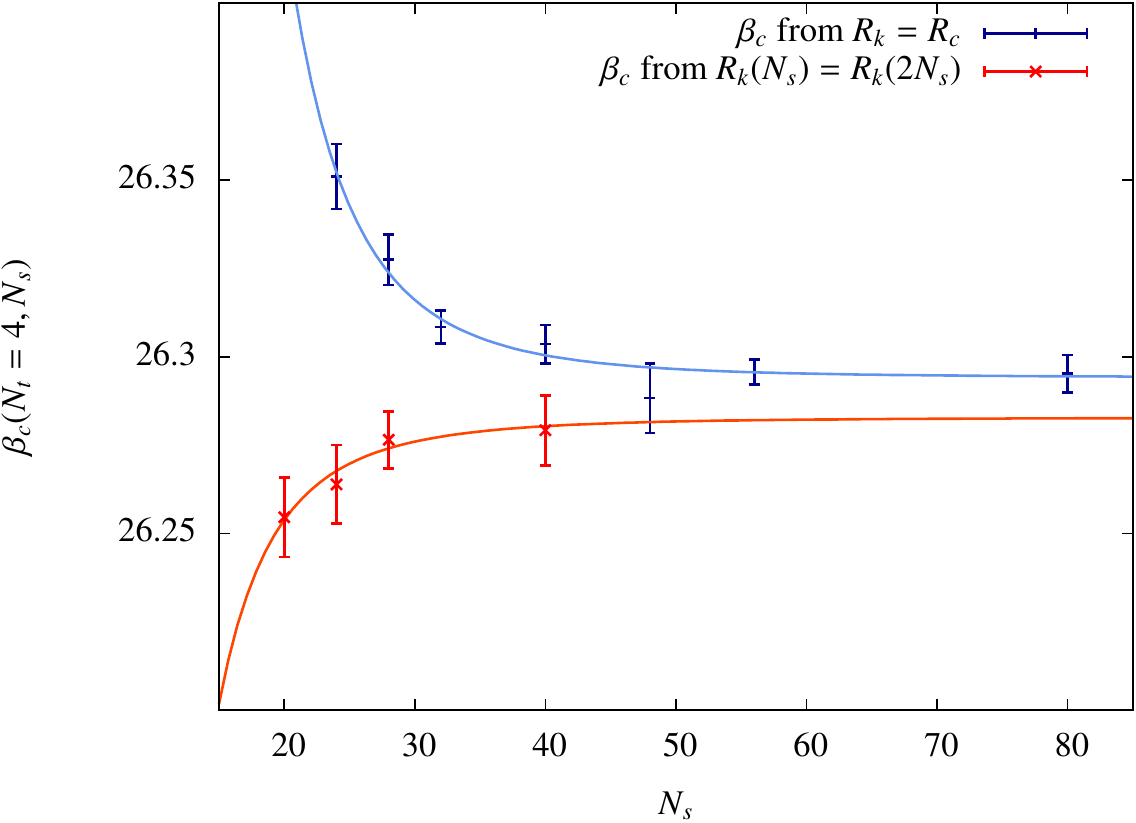}
\label{fig:su4crit}
}
\subfigure{
\includegraphics[width=0.47\textwidth]{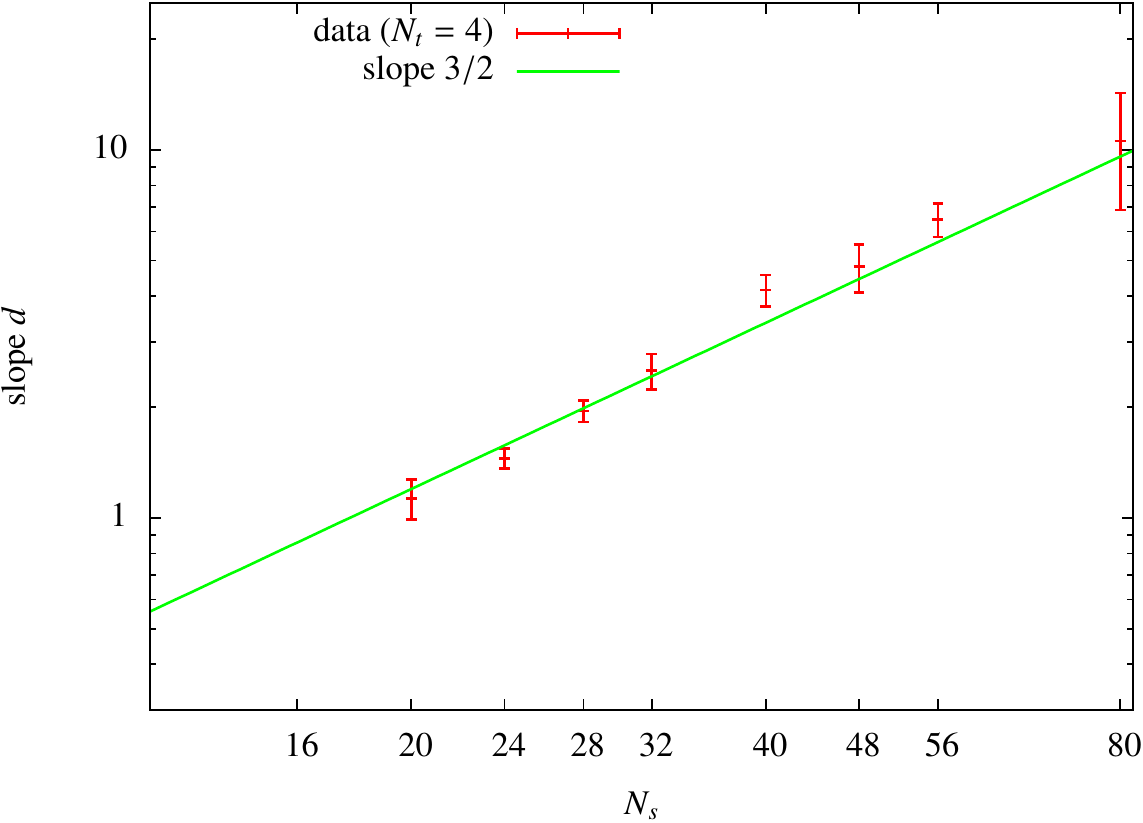}
\label{fig:su4nucheck}
}
\vspace{-.4cm}
\label{fig:su4plots}
\caption{SU(4) critical couplings (left), and
  slopes $d(N_s)$ compared to a power law with $\nu=2/3$ (right).}
\end{figure*}

\vspace{-.2cm}

\paragraph{Results for SU(4):} The present conclusion from a sequence of studies 
of the (2+1)$d$ SU(4) gauge theory 
\cite{deForcrand:2003wa,Liddle:2008kk,Holland:2007ar} is that the
deconfinement transition is weakly $1^\mathrm{st}$ order.
Especially the detailed analysis in \cite{Holland:2007ar} was
consistent with first order volume scaling laws.  
Here we assess to what extent Potts model scaling describes
the transition, at least approximately, and whether we find
indications of where our methods start to fail as we go to larger and 
larger volumes.

\begin{floatingfigure}[r]
\centering


\includegraphics[width=0.5\textwidth]{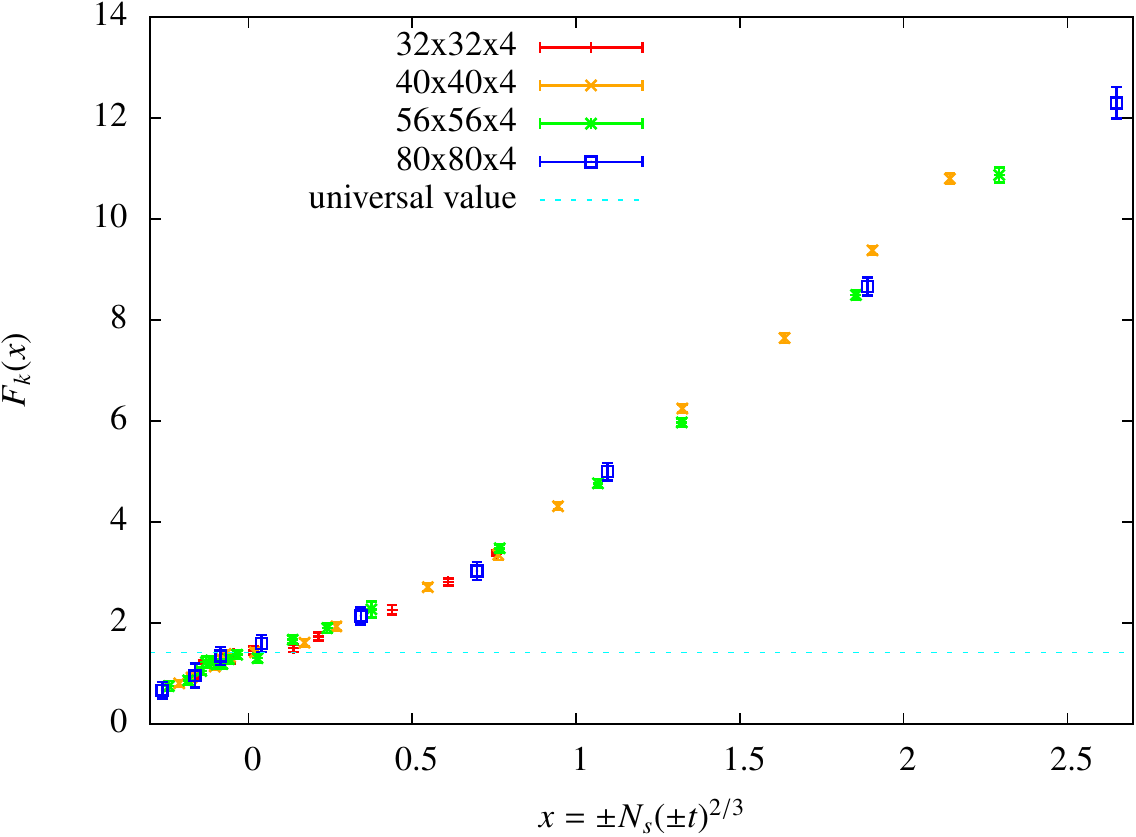}

\vspace{-.4cm}
\caption{ Check of Potts model scaling $(N_t=4)$}

\label{fig:su4scaling}
\end{floatingfigure}

First, we extract critical couplings from the pairwise intersections of
the $F_k$'s for pairs of lattices with $N_s$ ratios of 2:1. This method is
independent of Potts scaling and yields $\beta_{c}=26.283(9)$ for
$N_t=4$. Then we compare the so extrapolated
value to that obtained from intersecting the $F_k$'s with the 4-state Potts
universal value from \cite{PhysRevB.38.565}.  The latter has a
smaller error because we have more points to fit; it gives the 
consistent value $\beta_{c}=26.294(2)$. Pseudo-critical couplings
and fits for each method are shown in Fig.~\ref{fig:su4plots}.  
Both extrapolated values are consistent with $\beta_c=26.228(75)$ from
\cite{Liddle:2008kk} but deviate with some significance from
$\beta_c=26.251(16)$ given in \cite{Holland:2007ar}, where 
first order scaling was assumed in the infinite volume extrapolation
of the critical coupling.

\newpage

If we furthermore extract a critical exponent $\nu$ from the slopes
$d(N_s) $ of $F_k(\beta)$ at $\beta_{c}$ as before, see
Fig.~\ref{fig:su4plots}, we obtain  $\nu=0.60(2)$ from our data 
for $N_t=4$ with $\beta_{c}=26.283$. 
Some systematic uncertainty arises from what precise value is used
here, however.
Repeating the analysis for different values in the one-$\sigma$
interval around $\beta_{c}=26.283$, we find $\nu=0.59(5)$. 
Of all the $Z_4$-symmetric Ashkin-Teller models, which have continuous
$\nu \in [2/3, 1]$, this seems to be at best consistent with the lower
bound  $\nu=2/3$ for the $q=4$ Potts model, in agreement
with the earlier conclusion in \cite{deForcrand:2003wa}.  
It is a general trend of our method, observed also for SU(2) and
SU(3), that it underestimates the critical exponent
due to subleading finite-size effects, however.
Finally, our present $N_t=4$ data with spatial lattice sizes
up to $N_s=80$ shows reasonably good Potts scaling as seen in
Fig.~\ref{fig:su4scaling} where we plot the center-vortex free energy $F_k(x) $
over the scaling variable $x = \pm T_c L (\pm t)^\nu $ with $\nu = 2/3$. 

\vspace{-.4cm}

\section{Conclusions}

\vspace{-.2cm}

We have studied the deconfinement transition in the pure
SU(3) gauge theory in 2+1 dimensions on the lattice. Using 't
Hooft's twisted boundary conditions we have measured center-vortex
free energies and demonstrated that the self-duality of the
associated Potts model is directly reflected in SU(3): the free
energies of the confining electric fluxes are mirror images around
$T_c$ of those of spatial center vortices. We demonstrated how this can be
exploited to remove the leading finite-size corrections in the
determination of  critical couplings from numerical simulations. 
We do not yet have the data necessary to compute electric fluxes and to
test self-duality analogously in SU(4), but our available data does
not show any significant violations of the $q=4$ Potts scaling. 


\medskip

{
\noindent\textbf{Acknowledgements:} This work was supported by the
Helmholtz International Center for FAIR within the LOEWE program of 
the State of Hesse, the Helmholtz Association Grant VH-NG-332, and the
European Commission, FP7-PEOPLE-2009-RG No.~249203. Simulations were
performed on the high-performance computing facilities of eResearch
SA, South Australia.  }

\end{document}